\documentclass[prletter,showpacs,twocolumn,typeset,superscriptaddress]{revtex4}   
\usepackage{amsfonts}
\usepackage{amsmath}
\usepackage{amssymb}
\usepackage{graphicx}
\usepackage{epsfig}
\usepackage{citesort}

\usepackage{dcolumn}
\usepackage{bm}

\begin{document}

\preprint{APS/XXX-XXA}

\title{Quantum state stability against decoherence}
\author{A. Kr\"{u}gel}
\affiliation{Institut f\"{u}r Festk\"{o}rpertheorie, Westf\"{a}lische
Wilhelms-Universit\"{a}t, Wilhelm-Klemm-Str. 10, D-48149 M\"{u}nster,
Germany.}
\author{Luis Roa}
\affiliation{Center for Quantum Optics and Quantum Information, Departamento de F\'{\i}sica,
Universidad de Concepci\'{o}n, Casilla 160-C, Concepci\'{o}n,
Chile.}
\author{C. Saavedra}
\affiliation{Center for Quantum Optics and Quantum Information, Departamento de F\'{\i}sica,
Universidad de Concepci\'{o}n, Casilla 160-C, Concepci\'{o}n,
Chile.}

\date{\today}

\begin{abstract}
We study the stability of the coherence of a state of a quantum
system under the effect of an interaction with another quantum
system at short time. We find an expression for evaluating the order of magnitude
of the time scale for the onset of instability as a function
of the initial state of both involved systems and of the
sort of interaction between them. As an application we study
the spin-boson interaction in the dispersive interaction regime,
driven by a classical field. We find, for this model, that the
behavior of the time scale for the onset of instability, with
respect to the boson bath temperature, changes depending on the
intensity of the classical field.
\end{abstract}


\maketitle

\section{Introduction}

The question as to whether or not a pure quantum state can persist
in the macroscopic world has been considered since Schr\"{o}dinger
introduced his gedanken experiment known as Schr\"{o}dinger's cat
\cite{Schroedinger} and it was in this way that the entanglement
concept was introduced into the quantum world.
%
Decoherence is a word to indicate that the state of a quantum system is not pure.
In general, a hamiltonian interaction which generates entanglement between two quantum systems 
produces reversible decoherence in each of the involved systems.
However, when a system interacts with an infinite number of systems or a single system interacting with an infinite number of degrees of
freedom, the concept of \textit{irreversible decoherence} arises,
which is usually called \textit{decoherence} only.
Thus, \textit{decoherence} has become the terminology for the
irreversible evolution of a quantum state due to its
interaction with an environment \cite{Zurek}.
In this context numerous works allow us to gain intuition about the dynamical behavior of open quantum
systems.
For instance,
spontaneous emission arises from the coupling of a system to a
noise vacuum environment \cite{Weisskopf}, Dalvit \textit{et al.} \cite{Dalvit} studied various
measures of classicality of the \textit{pointer states} of open
quantum systems subjected to decoherence. Also J. I. Cirac
\textit{et al.} \cite{Cirac} found a \textit{dark state} of a single two-level
ion trapped in a harmonic potential by controlling its motion degree of
freedom. This is achieved by generating a squeezed motion
state. T. A. Costi and R. H. McKenzie \cite{Costi} gave a
quantitative description of the entanglement between a two-level
system and an environment for an ohmic coupling. D. DiVincenzo
and D. Loss \cite{DiVinchenzo} provided an exact analysis of the
weak coupling limit of the spin-boson model for an ohmic heat
bath in the low temperature limit, using non-Markovian and Born
approximations. S. Bose \textit{et al.} \cite{Bose} studied the
enforcer of entanglement between a two-level system and a
quantized mode in a thermal state. Decoherence-free
subspaces in cavity QED have been found \cite{Bosco}.
Also a method has been developed which achieves the slowing down of
decoherence and relaxation by fast frequency
modulation of the system-heat-bath coupling \cite{Agarwal}.
In addition A. G. Kofman and G. Kurizki \cite{Kofman} developed a unified theory of dynamically
suppressed decay and decoherence by an external field in qubits to arbitrary thermal bath and dephasing sources. 
The problem of stability of a quantum state under a class of Lindblad dissipative dynamics has been also studied
\cite{Retamal}.

In this article we study, in a simple form, the problem of
coherence stability, at short time, of a quantum state without Markoff nor
Born approximations.
We have found an expression which allows us to estimate the time scale
order of onsetting the decoherence for a
quantum system when it interacts only with another single quantum system.
The expression can also be applied when the second interacting system is in a thermal equilibrium or is in
an adiabatic dynamics regimen with respect to the studied one.
We apply our result to the following cases:
$(i)$ the reversible pure-dephasing interaction with a numerical simulation,
$(ii)$ a cavity mode driven by a thermal light, and
$(iii)$ the spin-boson dispersive interaction driven by a resonant classical
field.

\section{Decoherence rate}   \label{decoherence}

The stability against decoherence is understood to be the process where
quantum coherence is preserved along evolution \cite{Retamal}.
So, we can say that an initial pure state $|\psi\rangle$
is stable against decoherence during the time
$t_d$ if $tr\rho^2(t)\approx 1$ for all $t\le t_d$.
The decoherence of a state, represented by $\rho$, is measured by means
of the first-order entropy:
$s(t)=1-tr\rho^2(t)$.
In order to consider the stability of the coherence, we assume
that $s(t)$ is an analytic function of $t$ \cite{Retamal}.
To find out the $t_d$ stability time scale order for which $s(t)$ remains being
approximately $s(0)=0$, it can be expanded as a Taylor series,
in such a way that the time scale order will be given by
$
t_d=1/\sqrt[n]{s_n(0)},
$
where $s_n(0)$ denotes the $n$th derivative of $s(t)$ evaluated at $t=0$, and $s_n(0)$
is the lowest order derivative different from zero.
The first two derivatives of the first-order entropy are:
$
s_1=-2tr(\rho\dot{\rho})$ and $s_2=-2tr(\dot{\rho}^2+\rho\ddot{\rho}),
$
with $\dot{\rho}$ and $\ddot{\rho}$ denoting, respectively, the first
and the second derivatives of $\rho$ with respect to $t$ at time $t$.

Now let us suppose that a system under study, labeled by $a$, is interacting with the $R$ system
through the $V(t)$ hamiltonian.
We consider the whole $a-R$ system t be isolated or:
$R$ being in a thermal equilibrium or in an adiabatic dynamics regimen.
Initially the $a$ system is in
a pure state $|\psi\rangle$ and the $R$ system is in the state $\rho_R$.
In the whole tensorial product Hilbert space, $\mathcal{H}_a\otimes \mathcal{H}_R$,
the dynamics of the composite system state $\rho(t)$
is driven by the Master equation \cite{Carmichael} ($\hbar=1$):
\begin{equation}
\dot{\rho}(t)=-i[V(t),\rho(t)],
\label{me}
\end{equation}
whereas the dynamics of the partial density operator of the $a$
system, $\rho_{a}(t)=tr_R(\rho(t))$, is governed by
\begin{equation}
\dot{\rho}_a(t)=-itr_R[V(t),\rho(t)],  \label{mea}
\end{equation}
where $tr_R$ denotes the tracing up over the $R$ system.

To find out if the initial pure state $|\psi\rangle$ is stable
under the dynamics described by the (\ref{me}) and
(\ref{mea}) Eqs., and besides, to obtain the expression for the $t_d$, we
consider the change in  $s(t)$ by calculating its derivatives
at $t=0$.
Since we suppose initially the state of the
system $a$ to be pure, then $s(0)=0$ and $s_1(0)=0$; hence the
second derivative of the first-order entropy, which could be
different from zero, is given by
\begin{equation}
s_2(0)=2^2\langle\langle V[V,|\psi\rangle\langle\psi|]\rangle_R
-\langle V\rangle_R\langle [V,|\psi\rangle\langle\psi|]\rangle_R\rangle_a, \label{s2}
\end{equation}
where we have denoted the average on the system $a$ ($R$) by the
subindex $a$ ($R$), and they are taken at $t=0$, and $V=V(0)$. Thus,
when the $|\psi\rangle$ pure state does not commute with $V(0)$,
e.i. $[V,|\psi\rangle\langle\psi|]\neq0$, the value of
$t_d=1/\sqrt{s_2(0)}$ gives a time scale order for the onset of decoherence.
We can also see from Eq. (\ref{s2}) that, in principle, $t_d$ (\ref{s2}) will be a functional
of both the initial pure state of $a$ as the initial
state of the $R$ system, and is proportional to $1/g$.
It is important to point out that $s_2(0)$, Eq.
(\ref{s2}), is the positive defined operator-correlation in the $R$ system between the $V$
and $[V,|\psi\rangle\langle\psi|]$ operators, averaged on
$|\psi\rangle$.
When the $|\psi\rangle$ state commutes with $V(0)$,
one must calculate the third derivative of the
first-order entropy at $t=0$ in order to have the time scale
order.

Here it is worth emphasizing that $t_d$ was found using neither
Markovian nor Born approximations.
It is well to recall that the Markov approximation is a \textit{coarse grained} dynamics description, in the sense that the time scale
in which the system is observed is much longer that the characteristic correlation time of the reservoir.
Since, in that case, fine temporal structure can not be seen, $s_1(0)\neq 0$ in general.
Our description of the dynamics of the system corresponds to a time scale smaller than the correlation time of the reservoir.


By way of examples first let us consider familiar interaction
models which generate reversible decoherence, that is,
a far from the resonance interaction which is
known as simplest \textit{pure-dephasing} mechanisms
\cite{Leggett,Bose1,Mahan} described by the
$V=g(b+b^{\dagger})\sigma_z$ hamiltonians \cite{Bose1}, where $\sigma_z$ is the
\textit{z}-component of the $\mathbf{\sigma}$ spin-1/2 operator
with eigenstates $|0\rangle$ and $|1\rangle$. $b$ and $b^\dagger $
are the boson annihilation and creation operators respectively,
and $g$ gives account of the effective coupling
strength.
For this case the decoherence time scale order is given
by
\begin{equation}
t_d=\frac{1}{2g\sqrt{\langle(\Delta(b+b^{\dagger}))^2\rangle_R}\sqrt{1-\langle\psi|\sigma_z|\psi\rangle^2}}, \label{tdpd}
\end{equation}
where $\sqrt{\langle(\Delta(b+b^{\dagger}))^2\rangle_R}$ is the
root-mean-square deviation of the $b+b^{\dagger}$ boson
quadrature at $t=0$.

We can see that the initial coherence of the boson state plays an
important role, i.e., for a fixed $|\psi\rangle$ state, an initial
boson squeezed state \cite{Caves,Roa} causes a decoherence
time scale smaller than one caused by an initial Fock state or by a thermal
state with equal average boson number. In this effective model,
$|0\rangle$ and $|1\rangle$ states are affected
only by a phase and each one is stable under this pure-dephasing
mechanism. So, from (\ref{tdpd}) we can see that, for a fixed boson
state, the states on the equator of the Bloch sphere have a smaller
decoherence time scale than the one for states being near to the
poles. It is worth noting that, for this pure-dephasing
mechanism, any two states, $|\psi\rangle$ and its orthogonal
$|\psi_\perp\rangle$, have the same $t_d$ \cite{RO}.
For the Jaynes-Cummings resonance \cite{Jaynes} interaction model, one can
show that two states, $|\psi\rangle$ and its orthogonal
$|\psi_\perp\rangle$, have different $t_d$ \cite{RO}. 

\begin{figure} [t]
\includegraphics[angle=360,width=0.40\textwidth]{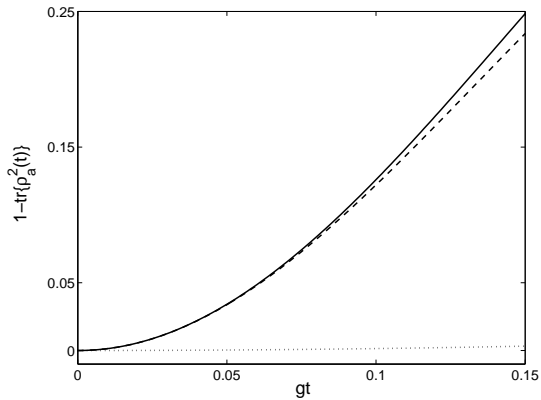}
\caption{Evolution of $s(t)$ for the two-level system under
an effective pure-dephasing interaction with a boson field mode.
The initial state of the boson mode is: a Fock state (solid), a
thermal state (dash), and a vacuum squeezed state with quantum
noise reduced in the $b+b^\dagger$ quadrature.} \label{figure01}
\end{figure}

Fig. \ref{figure01} shows the exact evolution of the $s(t)$
first-order entropy of the two-level system under an effective
pure-dephasing interaction with a boson field mode. The two-level
system is initially in the $|+\rangle$ eigenstate of $\sigma_x$
and the boson mode is in: a Fock state (solid), thermal state
(dash), and vacuum squeezed state (dot), each one with the same
average boson number $\langle n\rangle=3$. The $r$ squeeze
parameter is such that
$\sqrt{\langle(\Delta(b+b^{\dagger}))^2\rangle_R}=\sqrt{7-4\sqrt{3}}\approx
0.26795$ and for thermal and Fock state
$\sqrt{\langle(\Delta(b+b^{\dagger}))^2\rangle_R}=\sqrt{7}\approx
2.64575$.
Thus, in this particular case the $t_d$ time scale orders differ in one
order of magnitude as can be seen in Fig. \ref{figure01}.

As a second example, let us consider the explicit model which consists of
a cavity mode driven by a thermal light \cite{Carmichael,Retamal}.
In the Markov limit and Born approximation this model accounts for cavity losses.
The hamiltonian in the interaction picture of this physical model is given by
\begin{equation}
V(t)=\sum_j g_j(ar^\dagger_je^{i(\omega_j-\omega)t}+a^\dagger r_je^{-i(\omega_j-\omega)t}),
\end{equation}
where $\omega$ is the mode frequency and $\omega_j$ is the frequency of the $j$th mode of the thermal light.
$a$ and $a^\dagger$
are the annihilation and creation operators respectively, and $r_j$ and $r_j^\dagger $
are the annihilation and creation operators respectively of the $j$th mode of the thermal light.
$g_j$ account for the effective coupling strength between the main mode and the $j$th mode of the thermal light.
For this model the decoherence time scale is
\begin{equation}
t_{d}=\frac{1}{2\sqrt{(\gamma+2\gamma_T)\left( \langle a^{\dagger }a\rangle -\langle
a^{\dagger }\rangle \langle a\rangle \right) +\gamma_T}}, \label{td}
\end{equation}
being $\gamma=\sum_{j}\left\vert g_{j}\right\vert ^{2}$ which gives account of the whole magnitude of
effective couplings strength, $\gamma_T=\sum_{j}\left\vert
g_{j}\right\vert ^{2}\langle n_{j}\rangle$, and they are the reservoir correlation functions \cite{Carmichael}.
From Eq. (\ref{td}) we can see that, at zero temperature ($\langle n_j\rangle=0$ $\forall j$),
a coherent state is stable at this short time scale.
Retamal and Zagury \cite{Retamal}
show that this remains valid under the Markovian and Born approximations (Lindblad).
At finite temperature
a coherent state has a decoherence rate time scale given by
$1/\gamma_T$.
We also see that, for this irreversible dissipation mechanism only the initial field coherence, $\langle
a^\dagger\rangle\langle a\rangle$, allows to increase
the decoherence time scale.

\section{Spin-boson model}   \label{model}

Now let us consider a two-level system interacting with an
external laser mode of frequency $\omega_f$ and with a bosonic
bath modeled by an infinite collection of quantized harmonic
oscillators having frequencies $\omega_k$. The Hamiltonian which
drives the unitary dynamics of the whole system, in the
$H_{R}=\omega_f(\sigma_{z}+\sum_{k}b_{k}^{\dagger}b_{k})$ rotating
wave frame has the form ($\hbar=1$):
\begin{eqnarray}
H&=&\Delta\sigma_{z}+\sum_{k}\Delta_{k}b_{k}^{\dagger}b_{k}+\Omega
\left(  \sigma_{+}+\sigma_{-}\right) \nonumber \\
&&+\sum_{k}g_{k}\left(b_{k}
\sigma_{+}+b_{k}^{\dagger}\sigma_{-}\right), \label{H}
\end{eqnarray}
where $\Delta=\Delta_G-\omega_f$ and $\Delta_G=\omega_1-\omega_0$ represent the energy difference
between the upper state $|1\rangle$ and the lower state $|0\rangle$ of the two-level system.
Operators $\sigma_{z}$, $\sigma_{\pm}$, obey the standard SU($2$) commutation relations.
$\Delta_k=\omega_k-\omega_f$, and $g_{k}$'s are the respective coupling constants for the
dipolar interaction between the $k$th mode and the two-level system.
$\Omega$ stands for the Rabi frequency
which determines the coupling to the external classical field.
$b^{\dagger}_k$ and $b_k$ are the creation and the annihilation
bosonic operators of the $k$th mode, respectively.
We assume the external field to
be near resonance with the transition of the two-level system whereas the
boson modes are considered to be far off resonance in such a way that $\Delta_k\gg \omega_{k}$.

Therefore, boson mediated transitions, described by the fourth
term on the right hand side of (\ref{H}), can be strongly suppressed.
Thus, the effective Hamiltonian approximately describing the
interaction process can be obtained from the (\ref{H})
Hamiltonian by using the method of Lie rotations
\cite{Klimov,Sainz}, namely applying to the (\ref{H}) Hamiltonian
the unitary transformation:
$U=\exp[ \sum_{k}\varepsilon _{k}( b_{k}\sigma
_{+}-b_{k}^{\dagger }\sigma _{-})]$,
with $\epsilon _{k}=g_{j}/\Delta _{k}\ll 1$. Neglecting terms of order higher than
$\epsilon _{k}$ we obtain the following $H_{eff}$ effective Hamiltonian:
\begin{widetext}
\begin{equation}
H_{eff}=\Delta\sigma _{z}
+\sum_k\Delta _{k}b_{k}^{\dagger}b_{k}
+\Omega\left(\sigma_++\sigma_-\right)
+\sum_k\frac{g_{k}^{2}}{\Delta_G}\sigma_+\sigma_-
+2\left[
\sum_{k,k^{\prime }}\frac{g_{k}g_{k^{\prime }}}{\Delta_G}b_{k}^{\dagger }b_{k^{\prime }}
+\Omega\sum_k\frac{g_{k}}{\Delta_G}\left(b_k+b_k^{\dagger}\right)
\right] \sigma _{z},
\label{Heff}
\end{equation}
\end{widetext}
where we have considered that
$\Delta_G-\omega_{k}\approx\Delta_G$. Thus, off-resonant
transition terms have been eliminated and in their stead appear
three diagonal terms in the $\sigma_{z}$ representation: a
constant shift only for the upper level (a classical Stark shift);
a random phase shift term similar to that which is described in Ref.
\cite{Carmichael}; and a pure-dephasing term (that depends on the
intensity of the bath modes) which gives rise to a randomized phase
between the two levels, but what is striking about it is that here
the coupling constant includes the Rabi frequency of the field
whereas, in the standard spin-boson modes hamiltonian, an external field
is not coupled at all. Thus, from first principles, an
effective Hamiltonian for describing the dispersive interaction of a
two-level system with a bosonic reservoir is obtained.

The transformation of $H_{eff}$ into the interaction picture separates the motion generated by
\begin{equation}
H_0=\Delta\sigma_z+\sum_k\Delta_kb_k^{\dagger}b_k+\sum_k\frac{g_k^2}{\Delta_G}\sigma_+\sigma_-
\end{equation}
from the motion generated by the interaction term $\tilde{V}=H_{eff}-H_0$, so that
\begin{equation}
V=e^{iH_0t}\tilde{V}e^{-iH_0t}=\sigma(t)+\hat{B}(t)\sigma_z,
\label{V}
\end{equation}
where $\sigma(t)=\Omega(\sigma_+e^{i\Delta t}+\sigma_-e^{-i\Delta t})$, and
\begin{eqnarray}
\hat{B}(t)&=&2\Omega\sum_k\frac{g_k}{\Delta_G}\left(b_k^{\dagger}
e^{i\omega_kt}+b_ke^{-i\omega_kt}\right) \nonumber \\
&&+\sum_{k,k'}\frac{g_kg_{k'}}{\Delta_G}\left(b_k^{\dagger}b_{k'}
e^{i(\omega_k-\omega_{k'})t}-\bar{n}_k\delta_{k,k'}\right), \label{B}
\end{eqnarray}
with $\bar{n}_k$ being the average boson number of the $k$th mode.
Here we emphasize that the effective interaction
hamiltonian (\ref{V}) has the same form of the ohmic spin-boson model \cite{Leggett,DiVinchenzo}.

Considering Eqs. (\ref{s2}) and (\ref{V}) we obtain the
following $t_d$ time scale order for this model:
\begin{equation}
t_{d}=\frac{1}{2\sqrt{\langle \hat{B}^{2}\rangle _{R}\left( 1-\langle \psi
|\sigma _{z}|\psi \rangle ^{2}\right) }},
\end{equation}
where we have supposed that each boson mode is initially in a
thermal state at absolute temperature $T$. Under that condition
the operator (\ref{B}) satisfies $\langle \hat{B}\rangle _{R}=0$,
and
\begin{equation}
\langle \hat{B}^{2}\rangle _{R}=\frac{4\Omega^{2}}{\Delta _{G}^{2}}
\sum_{k}g_{k}^{2}(2\bar{n}_{k}+1)
+\sum_{k,k^{\prime}}\frac{g_{k}^{2}g_{k^{\prime}}^{2}}{\Delta _{G}^{2}}
\left( \bar{n}_{k^{\prime }}+1\right) \bar{n}
_{k}.    \nonumber
\end{equation}

For a fixed bath state the eigenstates of $\sigma_z$, that is, the
poles of the Bloch sphere, are stable under decoherence effects.
Meanwhile, the states on the equator of the Bloch sphere are
the less stable. If there are no spin-boson interactions, that
is $g_k=0$ $\forall k$, all the $|\psi\rangle$ states are stable
since the classical field drives a unitary evolution. The
classical field affects decreasing the $t_d$ time scale order
through an effective dispersive interaction similar to the
pure-dephasing mechanism. We can distinguish two regimes: the
first one appears at weak field regime, this is,
$\Omega\ll g_k$; the second one turns up at strong field
limit, $\Omega\gg g_k$. At the high $T$ temperature limit: the
strong field regime has associated a $t_d$ given by:
\begin{equation}
t_{d,\Omega\gg g_k}=\frac{\Delta _{G}}{4\Omega \sqrt{2kT\gamma \left( 1-\langle \psi
|\sigma _{z}|\psi \rangle ^{2}\right) }},                   \nonumber
\end{equation}
whereas in the weak classical field limit the $t_d$ becomes
\begin{equation}
t_{d,\Omega\ll g_k}=\frac{\Delta _{G}}{2kT\gamma \sqrt{1-\langle \psi |\sigma _{z}|\psi
\rangle ^{2}}},                   \nonumber
\end{equation}
where $\gamma =\sum_{k}g_{k}^{2}/\omega _{k}$. Here we obtain an
important difference for the $t_d$ behavior in the two regimes: in
the strong field limit $t_d$ is proportional to
$1/\sqrt{T}$; meanwhile, in the weak field regime $t_d$ is
proportional to $1/T$, both at high temperature limit.

\section{Conclusions} \label{conclusions}

In summary we have found a general expression which allows us to
estimate the decoherence time scale order for which the
instability of a quantum state is onset. This decoherence
time scale is a functional of the initial states of both interacting
subsystems, and of the kind of interaction between
them.
It is worth pointing out that the found decoherence
time scale is not symmetric with respect to the involved subsystems.

Specifically, for a spin-boson weak and dispersive interaction driven by a
classical field we have found diverse behaviors of $t_d$ with
respect to the absolute temperature, depending on the intensity of
the external classical field.

One physical system which for instance fits these conditions
would be a semiconductor quantum dot, where the uppermost valence
band and the lowest conduction band can be represented by the
ground and the excited eigenstates of $\sigma_z$. It is well known
that the energy of a phonon is smaller than the transition energy of a
two-level quantum dot system \cite{Krummheuer,AK}. Thus, the far from
resonance constraint that we have imposed on the spin-boson model would be well
satisfied in a semiconductor quantum dot.

Further studies could involve other natural interactions and other
atomic configurations as well as a resonant quantum driven field mode.
Besides, one could study a more general
model for the decoherence mechanism, which would include both dephasing
and dissipation, that is, considering two infinite sets of
modes, one set around the resonance and the other one far from
the resonance.

\begin{acknowledgments}
This work was partially supported by Grants FONDECyT No. 1030671,
Milenio ICM P02-49F. A. K. thanks DAAD.
The Authors thank A. Klimov for valuable discussions.
\end{acknowledgments}

\end{document}